\documentstyle[12pt]{article} 
\newcommand{\be}{\begin{equation}} 
\newcommand{\ee}{\end{equation}} 
\newcommand{\beq}{\begin{eqnarray}} 
\newcommand{\eeq}{\end{eqnarray}} 
 
\newcommand{\eps}{\varepsilon} 
\newcommand{\p}{\partial} 
 
\newcommand{\tE}{\lefteqn{\smash{\mathop{\vphantom{<}}\limits^{\;\sim}}}E} 
\newcommand{\tP}{\lefteqn{\smash{\mathop{\vphantom{<}}\limits^{\;\sim}}}P} 
\newcommand{\tQ}{\lefteqn{\smash{\mathop{\vphantom{<}}\limits^{\;\sim}}}Q} 
 
\newcommand{\Pt}{\lefteqn{\smash{\mathop{\vphantom{\Bigl(}}\limits_{\sim} 
\atop \ }}P} 
\newcommand{\Qt}{\lefteqn{\smash{\mathop{\vphantom{\Bigl(}}\limits_{\sim} 
\atop \ }}Q}

\newcommand{\tNn}{\lefteqn{\mathop{\vphantom{'}}\limits_{\sim}}{\cal N}} 
 
\newcommand{\R}{R} 
 
\newcommand{\SA}{{\cal A}} 
\newcommand{\SSA}{{\bf A}} 
 
\newcommand{\tPb}{{\tP_{\smash{(\im)}}}} 
 
\newcommand{\nd}{{\cal N}_D} 
\newcommand{\N}{{\cal N}} 
 
\newcommand{\im}{\beta} 
\newcommand{\Rb}{{\rm \bf R}} 
\newcommand{\Cb}{{\rm \bf C}} 
\newcommand{\Nat}{{\rm \bf N}} 
 
\newcommand{\G}{{\cal G}} 
\newcommand{\D}{{\cal D}} 
 
\newcommand{\Op}{\Delta_S} 
\begin{document} 
%
%
\title{ 
{\Large \bf  
On choice of connection in loop quantum gravity 
} 
} 
  
\author{ 
Serguei Alexandrov\thanks{e.mail: alexand@spht.saclay.cea.fr.  
Also at V.A.~Fock Department of Theoretical Physics, St.~Petersburg 
University, Russia}
}
 
\date{} 
 
\maketitle 

\vspace{-1cm}
\begin{center}  
\it Service de Physique Th\'eorique, C.E.A. - Saclay, 91191 Gif-sur-Yvette 
CEDEX, France\\ 
Laboratoire de Physique Th\'eorique de l'\'Ecole Normale 
Sup\'erieure, 24 rue Lhomond, 75231 Paris Cedex 05, France
\end{center} 
 
 
\
 
\begin{abstract} 
We investigate the quantum area operator in the loop approach based 
on the Lorentz covariant hamiltonian formulation of general 
relativity. We show that there exists a two-parameter family of 
Lorentz connections giving rise to Wilson lines which are  
eigenstates of the area operator. For each connection the area 
spectrum is evaluated. In particular, the results of the su(2) 
approach turn out to be included in the formalism. 
However, only one connection from the family  
is a spacetime connection ensuring that the 4d diffeomorphism invariance 
is preserved under quantization. It leads to the area spectrum 
independent of the Immirzi parameter. 
As a consequence, we conclude that the su(2) approach must be 
modified accordingly to the results obtained since  
it breaks one of the classical symmetries. 
\end{abstract}

%
\section{Introduction} 
%
 
Recently, a new Lorentz covariant approach to loop quantum gravity was 
proposed \cite{SA,AV}. In its framework the Immirzi parameter problem 
\cite{Imir} 
arising in the standard su(2) approach can be solved. Namely, it has 
been shown that i) the path integral \cite{SA}, ii) the area spectrum 
derived from a modified loop approach \cite{AV} are independent of the 
Immirzi parameter. 
 
The key point of the latter derivation was the possibility 
to define Wilson line operators by use of a connection different 
from the canonical one. This is a general ambiguity of the loop approach 
to quantum gravity independently on the gauge group used. 
In fact, in the case of the Lorentz gauge group 
we have to use this ambiguity 
since the canonical Lorentz connection leads to the area operator 
nondiagonal on the states created by the Wilson lines. 
Therefore, to obtain the area spectrum one should 
search for a shifted connection for that the Wilson lines would be 
eigenstates of the area. In principle, 
one should take into account all possible choices of the connection and 
investigate the arbitrariness arising in the spectrum. 
In \cite{AV} it has been shown that if we 
require that the shift vanishes on the constraint surface there is a 
unique Lorentz connection diagonalizing the area operator. The 
resulting spectrum does not depend on the Immirzi parameter $\im$ and 
it is given by the values of two Casimir operators 
\begin{equation} 
{\cal S}= 8\pi\hbar G \sqrt{C(so(3)) -C_1(so(3,1))}. \label{as} 
\end{equation} 
 
However, one can ask: what will we obtain if we do not impose the 
condition of coinciding of the canonical and shifted connections on 
the constraint surface? Is the found connection unique in this case? 
Or, if not, what spectra do another connections give? 
 
The present paper is aimed to answer all these questions. In the next 
section we show that, indeed, there exists a two-parameter family of 
Lorentz connections leading to the area operator diagonal on the 
Wilson lines.  In Sec. III we derive the corresponding spectra of the 
area operator. 
In Sec. IV we impose an additional physical condition on 
the connection to be used: it should be a spacetime connection, 
i.e. it should properly transform under all 4d diffeomorphisms. As a 
result, we arrive to a unique connection, which has been already found 
in the previous paper \cite{AV} and, accordingly, to the area spectrum 
(\ref{as}) independent of the Immirzi parameter.  
In Sec. V we find that the 
standard results of the su(2) approach (for review, see \cite{Rov-dif}) 
have a Lorentz covariant extension and are included in the present 
formalism.  We discuss 
the meaning of the results obtained in the previous section 
for the su(2) approach and argue 
that it breaks one of the symmetries of the classical theory. The last 
section concludes the paper with some remarks. In appendix A one can 
find the basic ingredients of the covariant hamiltonian formulation. 
Appendix B is devoted to the investigation of the time diffeomorphisms 
in the canonical approach. 
 
We use the following notations for indices. 
The indices 
$i,j,\dots$ from the middle of the alphabet label the space coordinates. 
The latin indices $a,b,\dots$ from the beginning of the alphabet 
are the $so(3)$ indices, whereas 
the capital letters $X,Y,\dots$ from the end of the alphabet are 
the $so(3,1)$ indices.

%
\section{Lorentz connections in the covariant approach} 
%
 
The aim of this section is to find all Lorentz connections $\SA_i^X$ such  
that the states created by the corresponding Wilson lines   
\be 
U_{\alpha}[\SA]={\cal P}\exp\left(\int_a^b dx^i \SA_i^X T_X\right) 
\label{wl} 
\ee 
are eigenstates of the area operator. $T_X$ is a Lorentz generator
in an irreducible representation of so(3,1). 
(For all details concerning 
the Lorentz covariant formalism we refer to the paper \cite{AV} 
and appendix A.) 

We suppose that, as in the su(2) case, the states forming the physical 
Hilbert space are some kind of spin network states such that
their links are associated with the Wilson lines (\ref{wl}). Then
our requirement means that these new ``spin networks'' should be eigenstates
of the area. This goes again along the usual approach.
 
The quantum area operator is defined in terms of the so called smeared 
triad operators in the following way  
\beq  
&{\cal S}=\lim\limits_{\rho \to \infty}\sum\limits_{n} \sqrt{g(S_n)}, \qquad 
g(\Sigma)=g^{XY}\tP_X(\Sigma)\tP_Y(\Sigma),& \label{areaop} \\ 
&\tP_X(\Sigma)=\int_{\Sigma} d^2\sigma \, n_i(\sigma) 
\tP_X^i(\sigma), &  
\label{smtriad} 
\eeq  
where we used a partition of a measured surface $S=\bigcup_n S_n$. 
$n_i=\eps_{ijk} \frac{\partial x^j}{\partial \sigma^1}  
\frac{\partial x^k}{\partial \sigma^2 }$ is the normal to the surface. 
The necessary condition for the operator (\ref{areaop})  
to be diagonal on the states (\ref{wl})  
requires that the commutator of the connection $\SA_i^X$ and the triad  
multiplet $\tP^j_Y$ is proportional to $\delta^j_i$ \cite{AV}. 
This gives the first condition on the connection to be found. 
Other conditions arise from the fact that to be a Lorentz connection 
it must transform in a proper way under the gauge as well as diffeomorphism 
transformations. Thus we arrive to the following list of requirements: 
\beq 
i) &\ & \{ \G(n),\SA^X_i\}_D= 
\partial_i n^X -f_{YZ}^X n^Y \SA^Z_i,  \label{gtr} \\ 
ii) &\ & \{ {\cal D}(\vec N), \SA_i^X \}_D= 
\SA_j^X\partial_i N^j+N^j\partial_j \SA_i^X, \label{difftr} \\ 
iii) &\ & \{ \SA^X_i,\tP_Y^j\}_D \sim \delta_i^j.  \label{newAP}  
\eeq

Let us find all quantities satisfying the conditions  
(\ref{gtr})-(\ref{newAP}). They can be constructed from both the  
initial canonical connection $A_i^X$ and the triad multiplets. 
First of all, by dimensional reasons we can restrict ourselves to  
the quantities which are linear in the connection $A_i^X$ or 
in the spatial derivative $\p_i$. Then one can show that it is impossible  
to construct a connection or a vector under the gauge transformations 
also satisfying the condition (\ref{difftr}) from terms of the 
second type only. This means that  
terms linear in the canonical connection uniquely define the shifted 
connection. (Pure triad terms commute with $\tP^j_Y$ and would not  
contribute to Eq. (\ref{newAP}).)  
 
If to write down all possible structures linear in $A_i^X$ one obtains  
a 12-parameter family of such terms. 
One can note that all of them satisfy Eq. (\ref{difftr}).  
Then the condition (\ref{newAP}) reduces the number of 
independent parameters to 4. 
However, the resulting quantity has anomalous contributions in the gauge 
transformations. To cancel them we should add some terms constructed 
only from the triad multiplets such that they themselves satisfy  
Eq. (\ref{difftr}). The cancellation turns out to be possible and 
imposes additional two restrictions on the parameters. 
As a result, we end up with a two-parameter family of Lorentz  
connections diagonalizing the area operator  
\beq 
\SA_i^X &=& 
A_i^X+\frac{1}{2}\left( (1+\frac{a}{\im})g^{XX'} 
- \frac{1}{\im}(1-b) \Pi^{XX'}\right) 
I_{(Q)X'}^T\frac{R_T^Z}{1+\frac{1}{\im^2}}f^Y_{ZW}\Pt_i^W \G_Y 
\label{famcon} \\ 
&+& 
(a g^{XX'} + b \Pi^{XX'}) \left( I_{(Q)X'}^R A_i^W   
-  I_{(Q)X'}^{R} f^{WYZ} \Qt_i^Y \p_l \tQ^l_Z 
+ \frac12 f^{WYZ} \tQ^k_{X'} \tQ^l_Y \Qt_i^Z \p_k \Qt_l^R  \right)\Pi_{WR}. 
\nonumber 
\eeq 
 
What can we say about these connections?  
First of all, one can calculate the following Dirac brackets which 
turn out to be extremely simple: 
\beq  
\{ \SA^X_i,\tP_Y^j\}_D&=&\delta_i^j \left( 
(1-b)\delta^X_{X'} +a\Pi^X_{X'} \right) I_{(P)Y}^{X'} , \label{AP2} \\ 
\{ \SA^X_i,\Pt_j^Y\}_D&=&-\left( (1-b)\delta^X_{X'} +a\Pi^X_{X'} \right) 
\Pt_j^{X'}\Pt_i^Y, \label{APt2} \\ 
\{ \SA^X_i,I_{(P)}^{YZ}\}_D&=&0. \label{AI2} 
\eeq 
The remarkable feature of these relations is that the projectors  
$I_{(P)}$ and $I_{(Q)}$ behave like $c$-numbers with the point of view of 
the bracket algebra.  
Due to this behavior from the Jacobi identity it follows  
\be 
\{ \{ \SA_i^X, \SA_j^Y\}_D , \tP^k_Z \}_D=0. \label{Jac} 
\ee 
This means that 
the Dirac bracket of two shifted connections does not depend on the connection 
itself and is a function of the triad multiplet and its derivatives only. 
Therefore, if we quantize the system replacing the Dirac brackets by 
the quantum commutators, 
there will be no ordering ambiguity in this relation 
since in the right hand side there are only commuting objects.

%
\section{Area spectrum} 
%
 
Let us consider the action of the area operator on a Wilson line  
(\ref{wl}) where the connection $\SA_i^X$ is taken from Eq. (\ref{famcon}). 
Due to the simplicity of the new quantization rules 
(\ref{AP2}) the action does not depend on the embedding and 
is given by a matrix operator: 
\beq 
{\cal S}\hat U_\alpha [\SA ] &=&\hbar \hat U_{\alpha_1}[\SA ] 
\sqrt{\Op} \hat U_{\alpha_2}[\SA ], \nonumber \\ 
\Op&=& \left( a^2 I_{(Q)}^{XY} - (1-b)^2 I_{(P)}^{XY} 
-a(1-b)\Pi^{XY} \right) T_X T_Y. 
\eeq 
One can show \cite{AV} 
\beq 
g^{XY}T_XT_Y &=&C_1(so(3,1)),\label{L1Cas} \\ 
\Pi^{XY}T_XT_Y&=&C_2(so(3,1)), \label{L2Cas} \\ 
I_{(Q)}^{XY}T_XT_Y&=&C(so(3)). \label{SCas} 
\eeq 
Thus, the quantity in the square root is nothing else but 
a linear combination of these Casimir operators: 
\begin{equation} 
\Op =(a^2 + (1-b)^2) C(so(3)) -(1-b)^2 C_1(so(3,1)) 
-a(1-b) C_2(so(3,1)). 
\label{oper} 
\end{equation} 
As a result, the area operator turns out to be diagonalizable 
and its spectrum is given by the eigenvalues of 
the Casimir operators (\ref{L1Cas})-(\ref{SCas}) \cite{Ruhl}: 
\beq 
C_1(so(3,1))&=&n^2+\mu^2-1 ,\label{L1sp} \\ 
C_2(so(3,1))&=&-2i n\mu , 
\qquad \qquad 2n\in \Nat, \ \mu \in \Cb,  \label{L2sp} \\ 
C(so(3))&=&j(j+1), \qquad (j-n)\in \Nat . \label{Ssp} 
\eeq 
 
Let us restrict ourselves to the unitary representations of the 
Lorentz group. 
It is natural since, due to the Plancherel theorem, they are sufficient
to span all square-integrable functions on the group \cite{sfLor}.
Therefore, we expect that only these representations will
appear in the basis elements of the Hilbert space which is still
to be found.
There are two series of such representations. 
The {\it principal series} is described by purely imaginary $\mu=i\rho, 
\ \rho \in \Rb$ and the {\it supplementary series} corresponds to the case 
of $n=0, \ \mu=\rho, \ -1<\rho<1$. 
Therefore, if we use the unitary representations to generate 
physical states, the area spectrum is given by 
\beq 
{\cal S}^{pr}&\sim& \hbar 
\sqrt{(a^2 + (1-b)^2) j(j+1) +(1-b)^2 (\rho^2 - n^2+1)-2a(1-b) n\rho}, 
\label{areasp2p} 
\\ 
{\cal S}^{sup}&\sim& \hbar 
\sqrt{(a^2 + (1-b)^2) j(j+1) +(1-b)^2 (1-\rho^2)}. 
\label{areasp2s} 
\eeq 
 
The remarkable property of the obtained spectra is that 
 due to the conditions $j\ge n$ and $|\rho|<1$ the expressions in the square 
root in Eqs. (\ref{areasp2p}) and  (\ref{areasp2s}) 
are strictly positive and the both spectra are always real. 
This is a different situation from that in Lorentzian 
spin foam models where an imaginary spectrum appears \cite{sfLor,Liv}. 
The reason what makes the area operator well defined can be traced 
out to the appearance of the Casimir of the SO(3) subgroup 
in the operator (\ref{oper}). It cancels the negative contribution 
from the Lorentz Casimir. Therefore, one can think that its 
presence is essential. And, indeed, it is impossible to obtain 
the spectrum given by the Casimirs of the Lorentz group only varying 
the parameters $a$ and $b$.

%
\section{Spacetime properties} 
%
 
We still have the two-parameter family of Lorentz 
connections (\ref{famcon}) and 
results for the area spectrum (\ref{oper}). 
Therefore, we should impose an additional physical condition to select the 
correct one. 
Our previous analysis was based on symmetries: we required that 
the symmetries of the classical theory are preserved under quantization. 
This requirement was expressed in terms of transformation 
properties of the connection used in the definition of the Wilson 
line. However, if we look at the conditions (\ref{gtr}) and (\ref{difftr}) 
we can see that one symmetry was missing so far. 
Namely, we did not consider transformation properties of our 
connection under the time diffeomorphisms. 
In fact, only the connection with the right transformation properties 
under all 4d diffeomorphisms can 
lead to quantum theory independent on the foliation, 
since only in this case the Wilson line can be generalized to the line 
in 4d space. 
 
Thus we arrive to the necessity to impose the following condition on 
the connections (\ref{famcon}): 
\beq 
iv)\ \delta_{(\xi^0)}\SA_i^X&=&\xi^0\p_0 \SA_i^X+\SA_0^X\p_i \xi^0, 
\label{trtime}\\ 
\delta_{(\xi^0)}\SA_0^X&=&\p_0(\xi^0\SA_0^X), \nonumber 
\eeq 
where $\SA_0^X$ is the corresponding generalization of the time 
component of the initial connection $A_0^i$. 
We will call the connections satisfying Eq. (\ref{trtime}) spacetime 
connections. 
 
First of all, one should find the generator of the time diffeomorphisms 
expressed in terms of the constraints. 
It is given by the full Hamiltonian: 
\be 
\D_0(\xi^0) = \int d^3 x\, \xi^0 \left( 
{\cal N}_{\G}^X \G_X+\nd^i \D_i+\tNn H\right), \label{gtime} 
\ee 
where 
\beq 
 {\cal N}_{\G}^X&=& A_0^X-N^i A_i^X, \\ 
\D_i&=&H_i+A_i^X\G_X=\p_j\left( A_i^X\tPb^j_X\right)-\tPb_X^j\p_i A_i^X. 
\eeq 
This expression was found in \cite{Reis} in the case of the Ashtekar 
gravity. In the given case it is still valid, but there appear 
additional subtleties. They are discussed in appendix B 
where the action of the generator (\ref{gtime}) is investigated. 
It is shown that on the surface of 
equations of motion and the Gauss constraint the family (\ref{famcon}) 
contains only one Lorentz spacetime connection. 
It is obtained at vanishing parameters $a$ and $b$ and 
it is given by the following expression: 
\be 
\SA_i^X=A_i^X + \frac{1}{2\left(1+\frac{1}{\im^2}\right)} 
\R^{X}_{S}I_{(Q)}^{ST}\R_T^Z f^Y_{ZW}\Pt_i^W \G_Y.  \label{spcon} 
\ee 
It is quite expected that it coincides with the initial canonical connection 
on the constraint surface since $A^X$ is definitely 
spacetime connection. 
From this it is clear that 
all other connections from the family (\ref{famcon}) 
are not spacetime since they do not coincide with $A^X$ on the 
constraint surface. 
 
Remarkably, the connection (\ref{spcon}) coincides with the one found 
in \cite{AV}. As a result, the corresponding spectrum of the area 
operator is given by Eq. (\ref{as}) 
and does not depend on the Immirzi parameter. 
 
The origin of this independence can be traced out to the independence 
of $\im$ of the commutator of the triad multiplet with the shifted 
connection 
\be 
[\SA_i^X,\tP_Y^j]=i\hbar \delta_i^j I_{(P)Y}^X. \label{APright} 
\ee 
This allows to hope that no dependence on the Immirzi parameter will appear 
in all physical quantities.

%
\section{Results for the su(2) approach} 
%

Note, that if we choose the parameters $a$ and $b$ in Eqs. 
(\ref{areasp2p}), (\ref{areasp2s}) 
to be as follows 
\begin{equation} 
a=-\im, \qquad b=1 
\label{parSU2} 
\end{equation} 
we reproduce the area spectrum found in loop quantum gravity 
\cite{area,ALarea} 
\begin{equation} 
{\cal S}\sim \hbar \im\sqrt{ j(j+1) }. 
\label{asSU2} 
\end{equation} 
Moreover, the connection used in this case is a Lorentz generalization 
of the standard su(2) connection. 
Indeed, from Eq. (\ref{famcon}) with (\ref{parSU2}) we find 
\beq 
\SSA_i^X &=& 
I_{(Q)Y}^X(\delta^Y_Z-{\im}\Pi^Y_Z) A_i^Z \nonumber \\ 
&-& \im \R^{XX'}\Pi_{WR} \left( 
 I_{(Q)X'}^{R} f^{WYZ} \Qt_i^Y \p_l \tQ^l_Z 
- \frac12 f^{WYZ} \tQ^k_{X'} \tQ^l_Y \Qt_i^Z \p_k \Qt_l^R  \right). 
\label{conSU2} 
\eeq 
It is easy to see that the second term 
is proportional to the field $\chi$ which vanishes in the time gauge 
and in the rest one can recognize the Ashtekar-Barbero connection \cite{barb} 
\be 
\SSA_i^X\mathop{=}\limits_{\chi=0} (0,\frac12 {\eps^a}_{bc}\omega_i^{bc}- 
\im\omega_i^{0a}). 
\ee 
Thus the standard results of the su(2) approach 
are included in the developed formalism. 
This fact allows to make some conclusions about the validity 
of the su(2) approach with the point of view of the covariant quantization. 
 
First of all, it is clear that to restore the full Lorentz symmetry we have 
to work with the Lorentz extension (\ref{conSU2}) of the Ashtekar-Barbero 
connection. Otherwise the symmetry is explicitly broken. 
The connection (\ref{conSU2}) possesses a remarkable property. In the time 
gauge it turns out to be commutative like its gauge fixed counterpart 
in the su(2) approach. Indeed, the first term in Eq. (\ref{conSU2}) 
can be rewritten as 
\be 
-\im\left(1+\frac{1}{\im^2}\right) 
 \Pi^X_Y I_{(P)Z}^Y(R^{-1})^Z_W \SA_i^W, \label{fterm} 
\ee 
where $\SA_i^X$ is the connection from Eq. (\ref{spcon}). 
One can show by explicit calculations that the quantity (\ref{fterm}) 
is commutative. 
Then due to Eq. (\ref{AI2}), 
which means  $\{ \SA_i^X, \chi^a\}_D=0$, 
the commutator $\{ \SSA_i^X, \SSA_j^Y\}_D$ 
is proportional to $\chi$ and vanishes in the time gauge. 
 
However, despite this attractive property the analysis of the previous section 
tells us that the quantization based on the connection (\ref{conSU2}) 
spoils the 4d diffeomorphism invariance\footnote{First, the observation 
that the Ashtekar-Barbero connection is not a spacetime connection has 
been made by J.~Samuel \cite{Sam}. This work is partially inspired by 
his ideas.}. 
The Immirzi parameter problem 
arising in this case can be viewed just as a reflection of the 
break of this symmetry. Therefore, 
the su(2) approach cannot be considered as a correct quantization of gravity. 
There is only one way in the framework of loop approach to quantize 
general relativity 
preserving all symmetries of the classical theory and 
it is given by Eqs. (\ref{spcon}) and (\ref{APright}).

%
\section{Conclusion} 
%
 
In this paper we analyzed arbitrariness in the loop quantization of 
general relativity in a manifestly Lorentz covariant formalism. 
This arbitrariness arises due to the possibility to use different 
connections in the definition of the Wilson line operators 
which are supposed to create physical states. 
Imposing the requirements that the used connection is a Lorentz connection 
and diagonalizes the area operator, we found a two-parameter family of such 
connections and the corresponding area spectra. 
Then we further restricted admissible quantities requiring 
them to be spacetime connections, i.e. to transform in a proper way 
under all 4d diffeomorphisms. This condition is sufficient to select 
a unique connection satisfying all requirements. We argue 
that the loop quantization should be based on the Wilson lines 
defined by this spacetime Lorentz connection, since only in this way 
one can preserve all classical symmetries. 
 
An important byproduct of our analysis is that the standard su(2) 
approach cannot represent a correct quantization of gravity. 
We can make such a conclusion since we found a Lorentz covariant 
extension of the results of this approach. 
In particular, the Ashtekar-Barbero 
connection can be extended to a Lorentz connection (\ref{conSU2}) 
which turns out 
to be in the mentioned two-parameter family. The corresponding 
area spectrum (\ref{asSU2}) coincides with the usual one. 
However, this 
connection is not a spacetime connection and breaks the diffeomorphism 
invariance in quantum theory. The appearance of the Immirzi parameter 
in the spectrum of the area operator is a reflection of this phenomenon. 
 
In this sense the spectrum (\ref{as}) we arrived from the found spacetime 
connection is remarkable. It cures all problems arising in different 
approaches to quantum gravity: 1) it does not suffer from 
the Immirzi parameter problem, 2) it is always real 
(see Eqs. (\ref{areasp2p}), (\ref{areasp2s})) 
in contrast to the results in Lorentzian spin foam models 
\cite{sfLor,Liv}, 3) even the trivial representation gives 
a nonzero value of area (cf. Eq. (\ref{asSU2})). 
 
We should emphasize that the presented analysis is essentially quantum 
despite of the lack of any information about the structure of the 
Hilbert space. 
It repeats step by step what has been done in \cite{ALarea} 
for the case of the su(2) approach. The only made assumption 
is that the Wilson lines correspond to physical states and they are 
eigenstates of the area operator. All other is a consequence 
of the requirement to preserve the diffeomorphism and local Lorentz 
symmetries under quantization. 
 
Of course, a lot of problems have to be solved yet. First of all, 
one should modify the Hilbert space structure found in the su(2) 
approach \cite{cyl} 
to the case of the Lorentz gauge group. 
May be the most important question at the moment is which representations
should be taken into account. The answer on this question will allow 
to conclude what is the final form of the area spectrum which is derived 
in terms of Casimir operators only. The related problem is whether 
the spectrum is discrete or continuous. And, of course, it is necessary
to find it to start the derivation of the black hole entropy in the 
new setup.
Nevertheless, the obtained results 
can give us a guess how the Hilbert space should look like.
Foe example, they imply that the Wilson lines (\ref{wl}) should be
somehow restricted to a definite representation of SO(3) subgroup.
Moreover, the restriction must do not break the Lorentz invariance.
All this can give important consequences for the structure of the 
resulting Hilbert space and it will be the subject of the subsequent work 
\cite{futur}. 
Also, it would be interesting to understand the role of the Gauss 
constraint in recovery of spacetime transformation properties 
from the canonical 
formulation. May be, the most interesting application this analysis 
can find in spin foam models which also claim for Lorentz covariant 
quantization of general relativity.

\section*{Acknowledgements} 
The author would like to thank R. Livine, 
V. Lyakhovsky, M. Reisenberger, C. Rovelli 
and D. Vassilevich for fruitful discussions.  The work has been 
supported in part by European network EUROGRID HPRN-CT-1999-00161.

\appendix 
 
%
\section{Covariant Hamiltonian formulation} 
%
 
In this appendix we give a short summary of 
the Lorentz covariant canonical formulation. 
Its complete description can be found in \cite{SA}. 
 
The canonical formulation is obtained by means of the $3+1$ 
decomposition: 
\be 
e^0=Ndt+\chi_a E_i^a dx^i \qquad e^a=E^a_idx^i+E^a_iN^idt 
\ee 
The field $\chi_a$ describes the deviation 
of the normal to the spacelike hypersurface $\{ t=0\}$ from 
the time direction. 
The usually used time gauge corresponds to $\chi=0$. 
 
The decomposed action is given by 
\beq 
S_{(\im)} &=&\int dt\, d^3 x (\tPb^i_X\partial_0 A^X_i 
+A_{0}^X \G_X+\nd^i H_i+\tNn H), \nonumber \\ 
\G_X&=&\partial_i \tPb^i_X +f_{XY}^Z A^Y_i \tPb^i_Z, \nonumber \\ 
H_i&=&-\tPb^j_X F_{ij}^X,  \label{Scov} \\ 
H&=&-\frac{1}{2\left(1+\frac{1}{\im^2}\right)} 
\tPb^i_X \tPb^j_Y f^{XY}_Z R^Z_W F_{ij}^W, \nonumber \\ 
F^X_{ij}&=&\partial_i A_j^X- 
\partial_j A_i^X+f_{YZ}^X A^Y_i A^Z_j, \nonumber 
\eeq 
where the following $3\times 6$ matrix fields are introduced: 
 \beq 
 & A_i^{ X}=(\omega^{0a}_i,\frac12{\eps^a}_{bc}\omega_i^{bc}) 
     &{\rm -\ connection\ multiplet}, \nonumber \\ 
&  \tP_X^{ i}=(\tE^i_a,{\eps_a}^{bc}\tE^i_b\chi_c) 
     &{\rm -\ first\ triad\ multiplet}, \nonumber \\ 
&  \tQ_X^{ i}=(-{\eps_a}^{bc}\tE^i_b\chi_c,\tE^i_a) 
     &{\rm -\ second\ triad\ multiplet},  \\ 
&  \tPb_X^i=\tP_X^i-\frac{1}{\im}\tQ_X^i 
     &{\rm -\ canonical\ triad\ multiplet},\nonumber \\ 
& X,Y,\dots=1,\dots, 6 & i,j,\dots=1,\dots, 3. \nonumber 
\eeq 
The fields $\tQ,\ \tP$ and $\tPb$ form multiplets in 
the adjoint representation of the Lorentz algebra and 
$A$ is the true Lorentz connection. 
The triad multiplets have a clear interpretation.
In the first triad multiplet one can recognize the spatial components
of the so called $B$-field, $B=e \wedge e$ (see, e.g. \cite{Liv}),
whereas the second triad multiplet is composed of its Hodge dual $\star B$.
The canonical multiplet is introduced since just this linear combination
of $\tQ$ and $\tP$ plays the role of the momentum conjugated to $A$.
It is clear that all triad multiplets
are related by simple numerical matrices: 
\beq 
\tP^i_X=\Pi_X^Y\tQ^i_Y,&\qquad & 
\Pi^{Y}_X =\left( 
\begin{array}{cc}
0&1 \\ -1&0
\end{array}
\right)\delta_a^b,   \label{p-q}     \\ 
\tP^i_X=\frac{\R_X^Y}{1+\frac{1}{\im^2}}\tPb^i_Y,&\qquad & 
\R^{Y}_X =\left( 
\begin{array}{cc} 
1& -\frac{1}{\im} \\ 
 \frac{1}{\im}&1 
\end{array} 
\right)\delta_a^b.     \label{pb-q} 
\eeq 
Therefore, the use of this or that multiplet is the matter of convenience.
The metric in Lorentz indices is given by the Killing form: 
\be 
g_{XY}=\frac{1}{4} f_{XZ_1}^{Z_2}f_{YZ_2}^{Z_1},\quad 
g^{XY}=(g^{-1})^{XY}, \quad 
g_{XY} =\left( 
\begin{array}{cc} 
\delta_{ab}&0 \\ 0&-\delta_{ab} 
\end{array} 
\right). 
\ee 
 
Besides the first class constraints $\G_X,\ H_i$ and $H$, 
in the theory (\ref{Scov}) there are 
the second class constraints: 
\beq 
 \phi^{ij}&=&\Pi^{XY}\tQ^i_X\tQ^j_Y, \nonumber \\ 
\psi^{ij}&=&2f^{XYZ}\tQ_X^{l}\tQ_Y^{\{ j}\partial_l \tQ_Z^{i\} } 
-2(\tQ\tQ)^{ ij }\tQ_Z^{l}A_l^Z+ 
2(\tQ\tQ)^{l\{i  }\tQ_Z^{j\}}A_l^Z=0, \label{psi} \\ 
(\tQ\tQ)^{ij}&=&g^{XY}\tQ_X^i\tQ_Y^j. \nonumber 
\eeq 
They give rise to a nontrivial Dirac bracket. 
This bracket has an important property: when one of its arguments 
is a first class constraint it coincides with 
the usual Poisson bracket. The only exception is the case of the 
Hamiltonian constraint. Then if the second argument depends on the 
connection there is no coincidence. 
 
The Dirac brackets of the canonical variables are 
\beq 
\{ \tPb_X^i,\tPb_Y^j\}_D&=&0, \nonumber \\ 
\{ A^X_i,\tPb_Y^j\}_D&=&\delta_i^j\delta^X_Y 
-\frac12 R^{XZ}\left(\tQ^j_Z\Qt_i^W+\delta^j_i I_{(Q)Z}^W 
\right)g_{WY}, \\ 
\{ A^X_i,A^Y_j\}_D& = & -\{A_i^X,\phi^{kl}\}(D_1^{-1})_{(kl)(mn)} 
\{\psi^{mn},A^Z_r\}\{\tPb^r_Z,A_j^Y\}_D \nonumber \\ 
&&-\{A_i^X,\tPb^r_Z\}_D\{A_r^Z,\psi^{mn}\}(D_1^{-1})_{(mn)(kl)} 
\{\phi^{kl},A^Y_j\},  \nonumber 
\eeq 
where we introduced the so called inverse triad multiplets $\Qt_i^X$ 
and projectors $I_{(Q)}^{XY}$ (see \cite{SA}).

%
\section{Time diffeomorphisms in canonical formulation} 
%
 
Before to investigate transformation properties of fields under 
the time diffeomorphisms, one should check whether the quantity (\ref{gtime}), 
indeed, generates these transformations. 
To be the generator, it should satisfy the algebra 
of the 4d diffeomorphisms ($\D=(\D_0,\D_i)$): 
\be 
[ \D(\xi),\D(\eta)] = -\D([\xi,\eta]).  \label{4ddiff} 
\ee 
This relation can be checked by using the transformation laws 
of the Lagrange multipliers $\N^{\alpha}=({\cal N}_{\G}^X,\nd^i,\tNn)$, 
\be 
\xi^{\im}\delta_{\im} \N^{\alpha}=\p_0 \xi^{\alpha} 
-C^{\alpha}_{\im\gamma}\xi^{\im}\N^{\gamma} 
\ee 
and the equations of motion 
\beq 
\p_0 A_i^X=\{ \D_0(1), A_i^X \}_D, \label{eqA} \\ 
\p_0 \tPb^i_X=\{ \D_0(1), \tPb^i_X \}_D. \label{eqP} 
\eeq 
Since the constraints $\G_{\alpha}=(\G_X,\D_i,H)$ form the usual algebra 
(coinciding, for example, with the constraint algebra of the Ashtekar 
gravity) the result of \cite{Reis} is still valid and the relation 
(\ref{4ddiff}) is fulfilled. 
 
At the same time, one can note that if we take another equivalent set of 
constraints ${\tilde \G}_{\alpha}=(\G_X,H_i,H)$ 
and the corresponding Lagrange multipliers 
${\tilde \N}^{\alpha}=(A_0^X,\nd^i,\tNn)$, we obtain an anomalous 
term in the algebra (\ref{4ddiff}) which is proportional to 
the square of the Gauss constraint. 
The reason is that due to the noncommutativity of the connection 
different choices of the Lagrange multipliers lead to inequivalent 
quantizations. Fixing a Lagrange multiplier we 
fix the quantity which has vanishing brackets with the canonical variables. 
So if its redefinition depends on a canonical variable, 
it can give inequivalent results. 
In the given case it does happen and the situation turns out to be 
quite different from that in the Ashtekar gravity, for example. 
 
Now it is straightforward to obtain transformation laws of any field. 
It is enough to note the simple formula: 
\beq 
\{ \D_0(\xi^0), \varphi(A,\tPb) \}_D &=& \xi^0 \p_0 \varphi + 
d_{\xi^0}[\D_0,\varphi], \\ 
d_{\xi^0}[\D_0,\varphi]&=& \{ \D_0(\xi^0), \varphi \}_D 
-\xi^0 \{ \D_0(1), \varphi \}_D. 
\eeq 
It means that only terms with derivatives of $\xi^0$ should be taken 
into account. Then one can obtain 
\beq 
\{ \D_0(\xi^0), A_i^X\}_D &=& \xi^0 \p_0 A_i^X + A_0^X \p_i \xi^0 + 
d_{\xi^0}[\tNn H,A_i^X], \\ 
d_{\xi^0}[\tNn H,A_i^X] &=& - \frac{\tNn \p_j \xi^0} 
{2\left( 1+\frac{1}{\im^2}\right)}R^X_S \left( \delta_i^j 
I_{(Q)}^{ST} +\Qt_i^T \tQ^j_W g^{WS} - \Qt_i^S \tQ^j_W g^{WT}\right) 
R^{TZ} \G_Z.  \label{Aanom} 
\eeq 
The appearance of the anomalous term (\ref{Aanom}) is little bit 
strange. One can try to cancel it by a redefinition of the Lagrange 
multiplier ${\cal N}_{\G}^X$ so that the Hamiltonian 
constraint get an additional term proportional to $\G_X$. 
Then according to the reasoning in the previous paragraph it could change 
transformation properties. 
However, it turns out that only the diagonal terms $\sim \delta_i^j 
I_{(Q)}^{ST}$ can be canceled in this way. 
Thus we conclude that the Gauss constraint should be used to recover 
the spacetime form of the time diffeomorphisms. 
 
Then it is clear that only the connection obtained at $a=b=0$ from 
the family (\ref{famcon}) satisfy the condition (\ref{trtime}). 
Indeed, the time diffeomorphisms of all other connections 
contain a lot of additional terms nonvanishing on the constraint 
surface. (They contain even terms with the second derivative of $\xi^0$.) 
We do not give explicit expressions since they are not used.


\begin{thebibliography}{99} 
 
 
\bibitem{SA} 
S.~Alexandrov, Class.\ Quantum\ Grav.\ {\bf 17}, 4255 (2000) 
[gr-qc/0005085]. 
 
\bibitem{AV} 
S.~Alexandrov and D.~Vassilevich,  
Phys.\ Rev.\ D {\bf 64}, 044023 (2001) 
[gr-qc/0103105]. 
 
 
\bibitem{Imir} 
G.~Immirzi, 
Nucl.\ Phys.\ Proc.\ Suppl.\ {\bf 57}, 65 (1997) 
[gr-qc/9701052]. 
 
 
\bibitem{Rov-dif} 
M.~Gaul and C.~Rovelli, 
{\it Loop Quantum Gravity and the Meaning of Diffeomorphism Invariance} 
Lectures given at the 35th Karpacz Winter School on Theoretical Physics: 
From Cosmology to Quantum Gravity (1999) [gr-qc/9910079]. 
 
 
\bibitem{Ruhl} 
W.~Ruhl, 
{\it The Lorentz Group and Harmonic Analysis} 
(WA Benjamin Inc, New York, 1970). 
 
 
\bibitem{sfLor} 
A.~Perez and C.~Rovelli, 
Phys.\ Rev.\ D {\bf 63}, 041501 (2001) 
[gr-qc/0009021]; 
3+1 spinfoam model of quantum gravity with spacelike and timelike 
 components, gr-qc/0011037. 
 
 
\bibitem{Liv} 
R.~E.~Livine and D.~Oriti, 
Barrett-Crane spin foam model from generalized BF type action for gravity, 
gr-qc/0104043. 
 
 
 
\bibitem{Reis} 
M.~Reisenberger, 
Nucl.\ Phys.\ B {\bf 457}, 643 (1995) [gr-qc/9505044]. 
 
 
 
\bibitem{area} 
C.~Rovelli and L.~Smolin, Nucl.\ Phys.\ {\bf B442}, 593 (1995); 
J.~Lewandowski, {\it The operators of quantum gravity} lecture at the 
workshop on canonical and quantum gravity (Warsaw, 1995); 
S.~Frittelli, L.~Lehner, and C.~Rovelli, Class\  Quantum\ Grav.\ 
{\bf 13}, 2921 (1996). 
 
\bibitem{ALarea} 
A.~Ashtekar and J.~Lewandowski, Class.\ Quantum\ Grav. {\bf 14}, 
A55 (1997). 
 
 
\bibitem{barb} 
J.~F.~Barbero, Phys.\ Rev.\ D {\bf 49}, 6935 (1994); 
Phys.\ Rev.\ D {\bf 51}, 5507 (1995); 
Phys.\ Rev.\ D {\bf 51}, 5498 (1995); 
Phys.\ Rev.\ D {\bf 54}, 1492 (1996). 
 
 
\bibitem{Sam} 
J.~Samuel, 
Class.\ Quantum\ Grav.\ {\bf 17}, L141 (2000) 
[gr-qc/0005095]; 
Phys.\ Rev.\ D {\bf 63}, 068501 (2001). 
 
 
\bibitem{cyl} 
A.~Ashtekar and J.~Lewandowski, {\it 
Representation theory of analytic holonomy 
$C^{\star}$ algebras}  in {\it Knots and quantum gravity} ed. J.~Baez 
(Oxford: Oxford University Press, 1994); 
J.\ Geom.\ Phys.\ {\bf 17}, 191 (1995); 
J.\ Math.\ Phys.\ {\bf 36}, 2170 (1995). 
 
 
\bibitem{futur} 
S.~Alexandrov, R.~Livine and D.~Vassilevich,  
in progress. 
 
\end{thebibliography}
\end{document}